\documentclass[10pt,twoside,twocolumn,a4paper]{IEEEtran}
\usepackage{amsmath,graphicx,url,times,amssymb}
\usepackage{color}
\usepackage[templates]{genealogytree}
\usepackage{float}
\usepackage{caption}
\title{MIR\MakeLowercase{a}G\MakeLowercase{e}: MULTICHANNEL DATABASE OF ROOM IMPULSE RESPONSES MEASURED ON HIGH-RESOLUTION CUBE-SHAPED GRID IN MULTIPLE ACOUSTIC CONDITIONS}

\author{Jaroslav \v{C}mejla$^{1}$,
      Tom\'{a}\v{s} Kounovsk\'{y}$^{1}$,
      Sharon Gannot$^{2}$,
      Zbyn\v{e}k Koldovsk\'{y}$^{1}$ and
      Pinchas Tandeitnik$^{2}$\vspace{0.1in} \\
      $^1$ Acoustic Signal Analysis and Processing Group,\\
      Faculty of Mechatronics, Informatics and Interdisciplinary Studies, 
      Technical University of Liberec,\\
      Studentsk\'{a} 2, 461 17 Liberec , Czech Republic.\\
      \{Jaroslav.Cmejla, Tomas.Kounovsky, Zbynek.Koldovsky\}@tul.cz\\
      $^2$ Faculty of Engineering, Bar-Ilan University, Ramat-Gan, 5290002, 
      Israel.\\
      \{Sharon.Gannot, Pinchas.Tandeitnik\}@biu.ac.il
      }

\begin{document}

\maketitle

\footnotetext{This work was partly supported by The Czech Science Foundation 
through Project No.~17-00902S and by the Erasmus+ KA~107 project 
No.~2017-1-CZ01-KA107-034883.}

\begin{abstract}
We introduce a database of multi-channel recordings performed in an acoustic lab with adjustable reverberation time. The recordings provide information about room impulse responses (RIR) for various positions of a loudspeaker. In particular, the main positions correspond to 4104 vertices of a cube-shaped dense grid within a $\text{46}\times\text{36}\times\text{32}$~cm volume. The database thus provides a tool for detailed analyses of beampatterns of spatial processing methods as well as for training and testing of mathematical models of the acoustic field.
\end{abstract}

%\begin{keywords}
%Room Impulse Response, Acoustic Transfer Function, Microphone Array, Database, 
%Grid
%\end{keywords}

\section{Introduction}
\label{sec:intro}

An exact mathematical description of the sound propagation in acoustic environments is difficult to define, as this is influenced by the shape of the room and by all objects, materials, and other physical properties of the enclosure. Since the propagation is linear, it is characterized by room impulse responses (RIR) or, equivalently, acoustic transfer functions (ATF) in the frequency domain. A RIR characterizes the sound propagation from one location to another. %and depends also on the directivity and rotation of the source and microphone. 

When the RIRs relating some source positions and microphone array arrangements in a given room are known, the microphone signals can be generated with a high precision by filtering the emitted acoustic sources by the respective RIRs. This enables us to test signal processing algorithms and their behaviour in  real-world conditions and, in particular, to verify the spatial robustness of multi-channel processors, to analyze acoustic fields and beampatterns, etc.

%acoustic source as it is observed by a microphone, 
For simulating RIRs, it is popular to use generators based on the image method \cite{allen1979}, which provide easy testing of algorithms as how they respond to changes of simple parameters such as reverberation time T$_{60}$, reflection coefficient, source distance or array geometry. However, the image method assumes artificial rectangular rooms (`shoebox') with simple walls containing no objects and no diffusive materials. The generated RIRs thus correspond to linear combinations of fractional-delay filters and fail to exactly imitate RIRs of real rooms with similar shapes. 

Realistic RIRs can be obtained from RIR databases that were measured in real conditions;\footnote{\tt https://signalprocessingsociety.org/get-invol ved/audio-and-acoustic-signal-processing/online- resources} 
see e.g., \cite{jeub2009,kayser2009,steward2010,hadad2014}. Although various settings are typically considered, the user is always limited to the arrangements realized by the authors of the given database. For example, the database by \cite{hadad2014} considers three geometries of an 8-microphone linear array, with a source 1 and 2 meter from the array center at 13 angles from -90$^{\circ}$ to 90$^{\circ}$, and T$_{60}$ either 160ms, 360ms or 610ms. 

Other arrangements are useful for specific case studies, however, they are less useful for detailed analyses of spatial processors, e.g., their robustness to small perturbations in parameters \cite{zheng2004,barnov2018}. To analyze such details, a database containing dense measurements of RIRs must be realized, which is a cumbersome task as the measurements must be precisely repeated for different settings. Smaller databases of this kind were realized, e.g., in \cite{fakhry2012,koldovsky2013} to support specific analysis. Alternatively, dense sets of RIRs for various microphone positions can be obtained using large arrays; see, e.g., \cite{weinstein2007}.

This paper aims at filling this gap. A new database comprising dense measurements of RIRs is described. The measurements were realized in the acoustic lab at the Bar Ilan University using a device for precise positioning of a loudspeaker, allowing for positioning the source positions in a dense grid of points within a relatively small volume. Recordings were acquired using six linear arrays of microphones; the measurements were repeated for three different levels of reverberation time.
The database provides a new tool for detailed analyses of spatial processing methods, e.g., for source enhancement, localization, and separation. Also, it can be used for learning and testing of new mathematical models of acoustics; field or of RIRs; see, e.g., \cite{talmon2013relative,bracha2015,bracha2016,katzberg2018}. The database, dubbed \mbox{MIRaGe}: Multichannel room Impulse Response database on Grid, will be available to the research community free-of-charge.
                                                                                                                                                                                                                                                                                                                                                                                                                                                                                                                                                                                                                                                                                                                                                                                                                                                                                                                                                                                                                                                                                                                                                                                                                                                                                                                                                                                                                                                                                                                                                                                                                                                                                                                                                                                                                                
%In general, typical RIRs have that many parameters as is their time-domain length. However, in a given room, they actually depend on fewer parameters such as source and microphone locations, rotations, etc. 

%However, a complete set of RIRs for all possible arrangements in the given room would be prohibitively large. Therefore, confined 2D or 3D areas are considered. 

The paper is organized as follows. Section~2 describes the recording setup, the resulting database, and the associated software package that will also be provided. Section~3 describes simple experiments, demonstrating the usage of the database. Section~4 concludes the paper.

\section{Database description}
% Firstly, we define the model of the recording setup then we provide detail of the realization and last we describe the contents of the finished database.

% The database contains sound recordings reproduced from various positions in the room that were recorded by six static linear arrays with 5 microphones each
\subsection{Setup}

\begin{figure}[h]
    \includegraphics[width=\columnwidth]{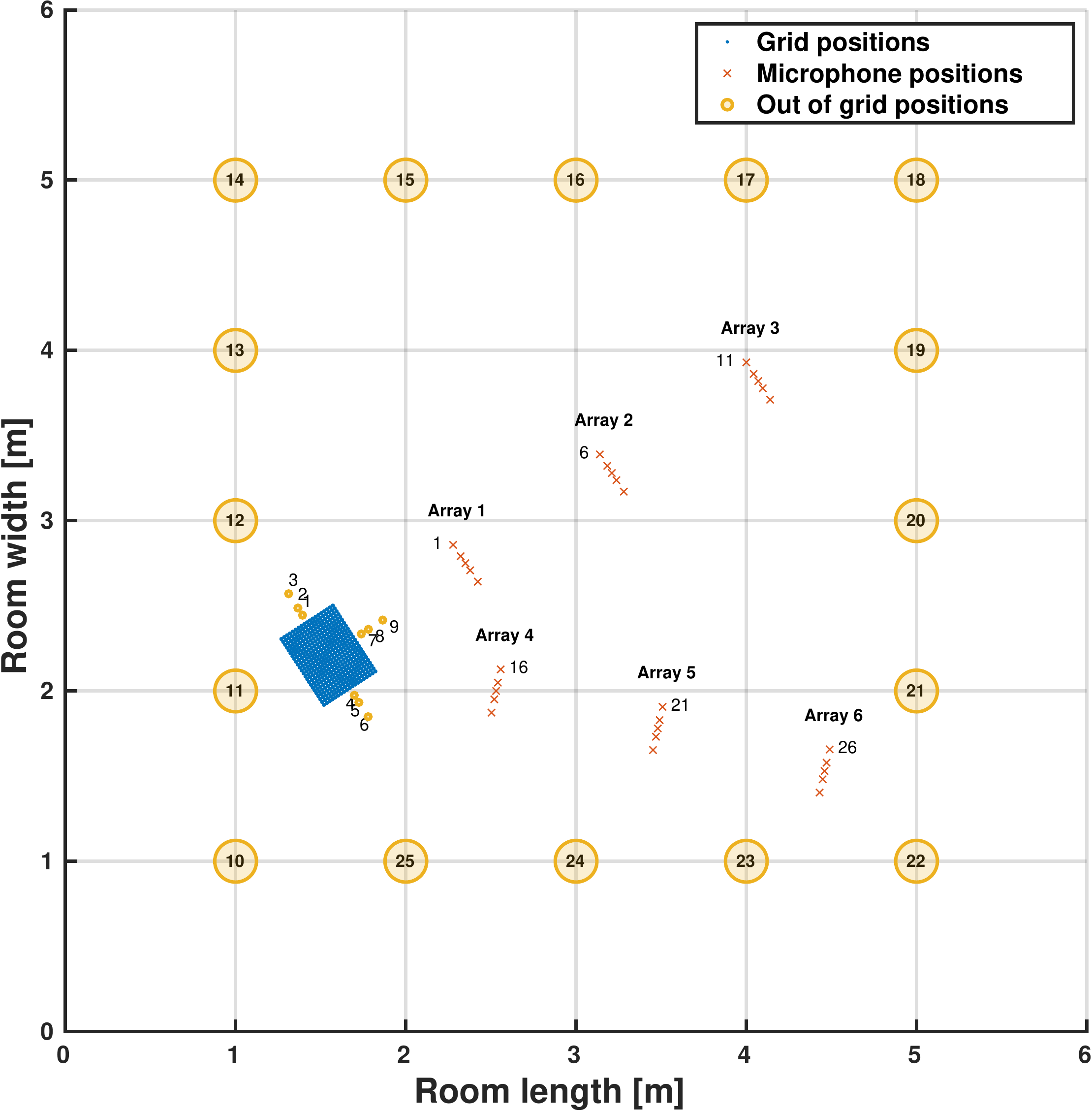}
    \caption{Illustration of the recording setup - Top View.}
    \label{fig:model-top}
\end{figure}

\begin{figure}[h]
  \includegraphics[width=\columnwidth]{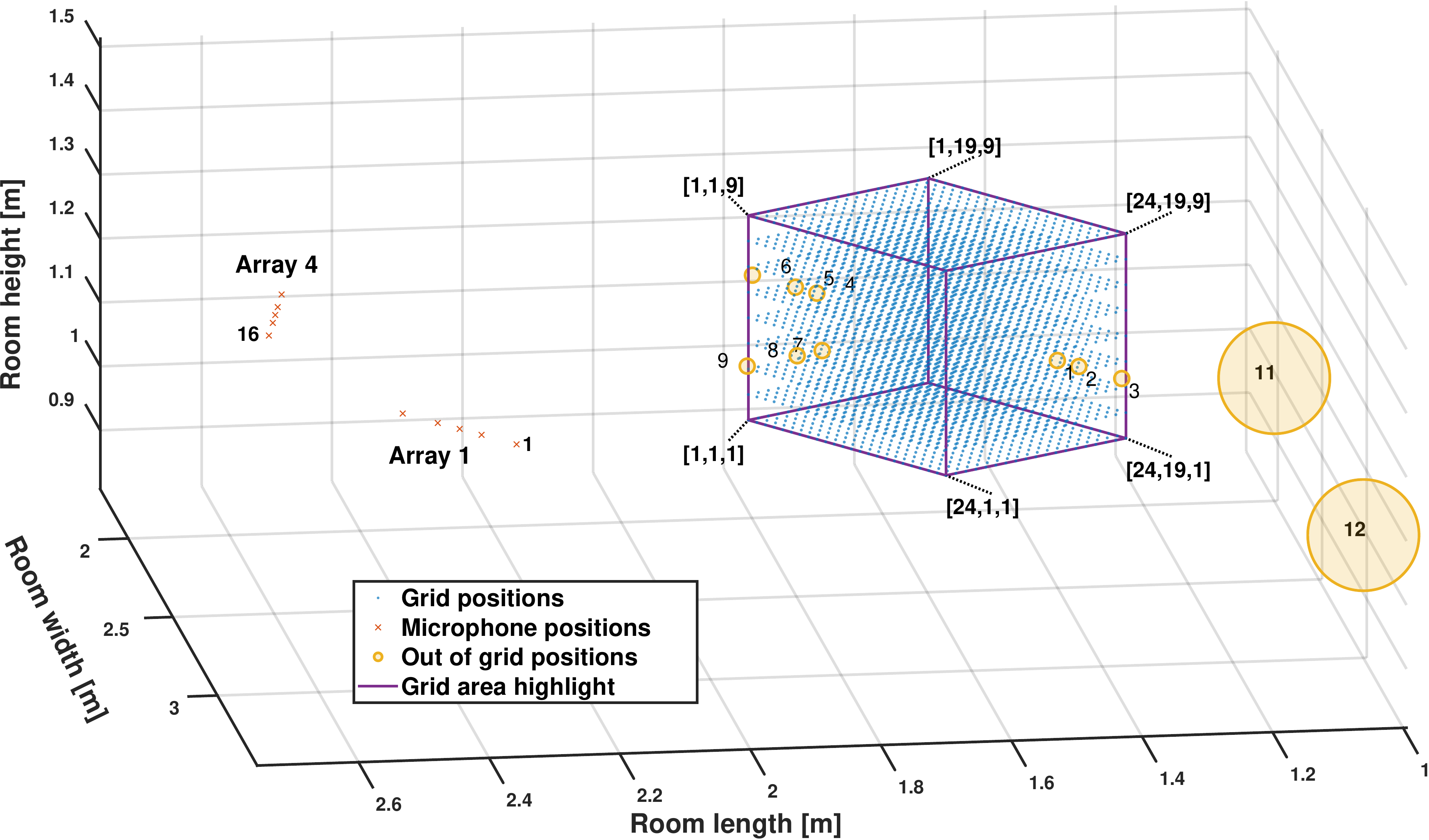}
  \caption{A detailed view of the near area of the grid.}
  \label{fig:model-detail}
\end{figure}

The recording setup is situated in the acoustic laboratory, which is a $6\times 6\times 2.4$~m rectangular room. A loudspeaker emulating the source is located within a cube-shaped volume of dimensions $\text{46}\times\text{36}\times\text{32}$~cm (from now referred to as {\em grid}) as shown in Figures \ref{fig:model-top} and \ref{fig:model-detail}. The positions of the loudspeaker form a grid sampled every $2$~cm across the $x$-axis and $y$-axis and every $4$~cm across the $z$-axis, so there are $\text{24}\times\text{19}\times\text{9}=4104$ possible source positions (grid vertices) in total. In the following, we will use the Matlab-like notation, for example, [:,:,1] means all vertices in the first horizontal slice of the grid.

In addition, there are $25$ source positions located outside of the grid (OOG); nine positions are close to the grid and $16$ positions are situated along the walls. The center of the grid (the 5th level), as well as the OOG positions, were positioned at the same height of $115$~cm. In the grid positions, the loudspeaker is directed in parallel to the $y$-axis towards the opposite corner of the room (referred to as `the front of the grid'). In the OOG positions, the loudspeaker is directed towards the center of the room.

The entire setup is recorded by six static linear microphone arrays and one microphone mounted $2$~cm in front of the loudspeaker, which is changing its position with the loudspeaker (microphone 31). Three microphone arrays are placed directly in front of the grid at the distance of $1$, $2$, and $3$~m from the center of the grid. The other three arrays are located at the angle of -45$^{\circ}$ at the same distances; see Fig.~\ref{fig:model-top}. All arrays are directed towards the grid centre %INSsg{broadside configuration indeed but not exactly braodside}
and placed at the same height of $115$~cm. Each array consists of $5$ microphones with the  inter-microphone spacing of $-13$, $-5$, $0$, $+5$ and $+13$~cm relative to the central microphone. %In addition, a single microphone was attached 2cm in front of the source/loudspeaker. The microphone is not static but rather changes position with the source/loudspeaker.

\begin{figure*}[ht!]
    \includegraphics[width=\textwidth]{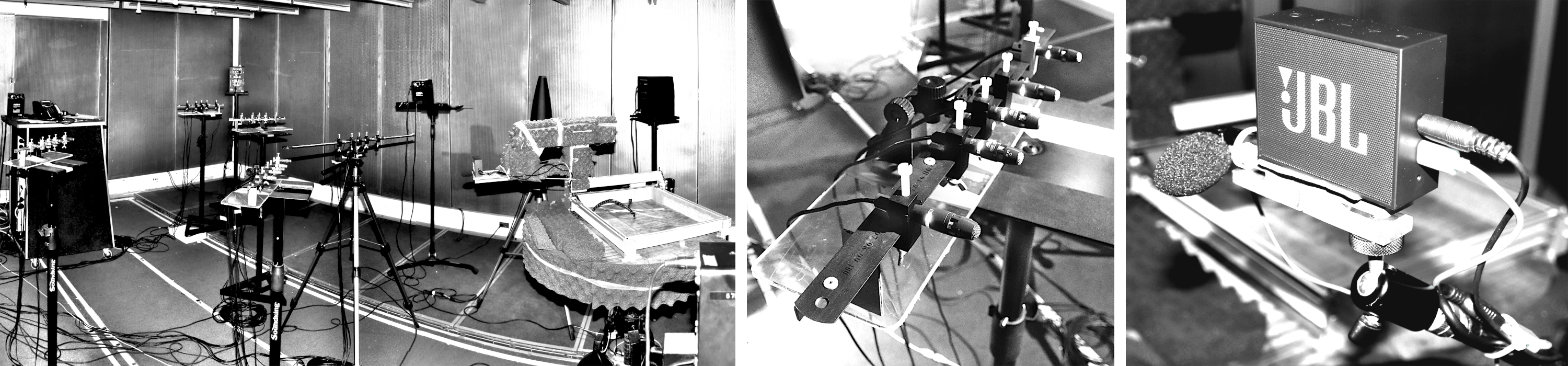}
    \caption{Photos of the recording setup in the acoustic lab. The first photo from left captures the whole room setup with the 3D positioning system and microphones arrays. The second photo shows the microphones array geometry. The third photo shows the loudspeaker located in an OOG position with the mounted microphone.}
    \label{fig:photos}
\end{figure*}

\subsection{Realization}\label{sec:realization}

\begin{figure}[H]
    \includegraphics[width=\columnwidth]{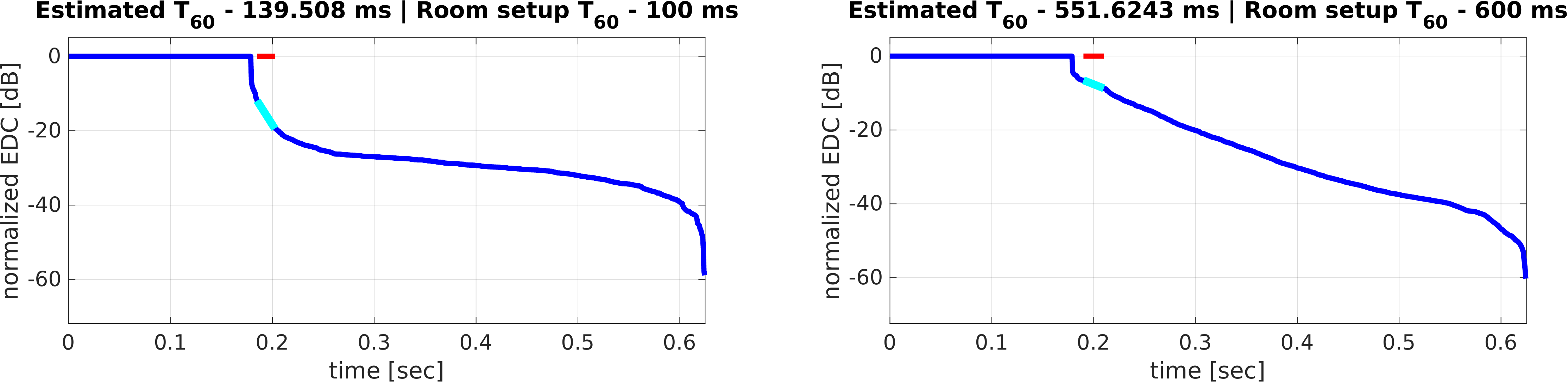}
    \caption{Samples of RIR Energy Decay Curves (EDCs) created from the MIRaGe database (mic 1, position [12,10,5]); $T_{60}$ was estimated from manually selected (marked - red, cyan) linear parts of EDCs}
    \label{fig:EDC_samples}
\end{figure}

%The database was recorded in a special acoustic room at BIU.  
The sidewalls and the ceiling of the acoustic lab consist of revolving double-sided panels with a reflecting and an absorbing face. By rotating these panels, the ratio between the reflective and absorbing areas of the room can be modified, by which the reverberation time of the room can be controlled. We have chosen three reverberation time levels: $100$, $300$, and $600$~ms; $T_{60}$ was measured by Br{\"u}el \& Kj{\ae}r type 2250 sound level meter employing Br{\"u}el \& Kj{\ae}r Omni source loudspeaker type 4295; see examples of RIR Energy Decay Curves in Fig.~\ref{fig:EDC_samples}. The other details of the recording hardware are summarized in Table~\ref{tab:rec_hw}.

The positioning of the loudspeaker within the grid was realized using a precise three-axis positioning system. It consists of a 2D plotter and a lift table. The $x$ and $y$-axis were controlled automatically by the plotter, while the $z$-axis was operated manually. The height was measured by a laser distance meter mounted to the desk of the table. To attenuate  acoustic reflections from the plotter's rails, a cardboard construction that holds the loudspeaker out of the plotter's perimeter was constructed (see the first photo in Fig.~\ref{fig:photos}).
Positions $10$ through $25$ were measured manually, so the accuracy of their positioning is slightly lower compared that of the positions in the grid.

For each position, two excitation signals were played and recorded in sequence: The first signal (Chirp) consists of two repetitions of a logarithmic swept-frequency cosine signal with a total length of $20$ seconds ($0.5$~s silence, $8$~s chirp, $2$~s silence, $8$~s chirp, $1.5$~s silence). White noise (WN) was used as the second excitation signal with a total length of $10$ seconds ($0.5$s silence, $8$s WN, $1.5$s silence). Due to the limited frequency range of the loudspeaker ($180$~Hz - $20$~kHz), the Chirp signals starts from $200$~Hz through $16$~kHz. To prevent rapid phase changes and subsequent ``popping", all excitation signals were linearly faded in and out over 0.2 seconds.

We have also recorded one hour of {\em room tone} (silence) and one hour of diffuse babble noise for each T$_{60}$ setting. The babble noise was simulated by using eight loudspeakers each playing a different multi-speech sequence and each placed approximately $1$~m from the walls: one in each corner and one in the middle of each wall. The loudspeakers were directed towards the nearest corner or the nearest wall. Four of these loudspeakers can be seen in the first photo in Fig.~\ref{fig:photos}. 

\begin{table}[t]
\caption{Recording equipment}
\label{tab:rec_hw}
\centering
\begin{tabular}{ll}
Microphones                   & AKG CK32      \\
Mic preamp. $+$ AD/DA         & ANDIAMO.MC    \\
Loudspeaker (grid, OOG)       & JBL Go        \\
Loudspeaker (babble noise)  & 6301bx Fostex
\end{tabular}
\end{table}

The recording/playback was done with the sampling frequency of $48$~kHz and the bit rate of $32$~bits per sample. In order to reduce the size of the raw database ($1.5$ TB), all recordings were re-encoded into the FLAC format (Free Lossless Audio Codec) with 48~kHz and 24~bits per sample (50\% reduction).

\begin{figure*}[ht]
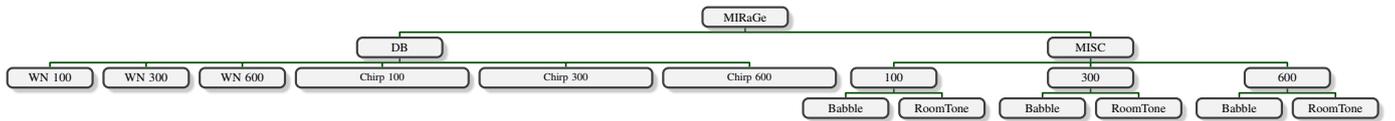

\noindent\resizebox{\textwidth}{!}{
\begin{genealogypicture}[
  template=signpost,
  level distance=2mm,
  level size=5mm,
  ]
child{
  g{MIRaGe}
  child{ g{DB}
    c{WN 100} c{WN 300} c{WN 600} c{Chirp 100} c{Chirp 300} c{Chirp 600}
  }
  child{ g{MISC}
    child{ g{100}
      c{Babble} c{RoomTone}
    }
    child{ g{300}
      c[id=node_C]{Babble} c{RoomTone}
    }
    child{ g{600}
      c[id=node_A]{Babble} c{RoomTone}
    }
  }
}
\end{genealogypicture}
}
\caption{Database directory structure: The acronyms WN and Chirp correspond to the excitation signals as defined in Section~\ref{sec:realization}. The numbers $100$, $300$, and $600$ correspond to the reverberation time level. MISC contains the recordings of the room tone and of the babble noise.}
\label{fig:folder_structure}
\end{figure*}

\subsection{Database package}
The database can be 
downloaded\footnote{\label{fn:link}\url{https://asap.ite.tul.cz/downloads/MIRaGe}}
 part-by-part according to the directory structure shown in 
Fig.~\ref{fig:folder_structure}. Each directory at the bottom level in 
Fig.~\ref{fig:folder_structure} comprises seven folders: one per each 
microphone array (01, \dots, 06) and one for the microphone mounted to the 
loudspeaker (on\_SPK\_mic). A software package for Matlab is available at the 
same webpage; the software documentation is 
included. 
It enables users to compute RIRs and relative RIRs or, equivalently, ATFs or 
Relative Transfer Functions (RTF) \cite{gannot2001} directly from the raw 
recordings for selected positions, microphones and the other parameters. The 
entire database has $0.7$~TB (FLAC) in size, however, only selected parts can 
be downloaded and used with the software.

%We are unable to provide any precomputed ATFs or Relative Transfer Functions (RTFs) due to the nigh-unlimited number of combinations of parameters (sampling frequency, length of the RIRs, reference microphone, etc). However, along with the database, we do provide a tool for simple ATF/RTF calculation \footnote{GitHub link}. 

% \subsection{Software utilities}
% Table of Matlab main scripts and functions, GUI?

% \subsection{Baseline dataset}
% pseudocodes - how we compute RIRs (parameter settings, resampling, least-squares, regularization), figures with some analysis

% The least-square method to compute the RIRs using the Levinson-Durbin algorithm with Tikhonov regularization. Selection of the Tikhonov regularization parameters (why we need this?). Any other details.

\section{Using the database: Examples} %Robust Beamforming for Noise Reduction
%We consider two experiments using the MIRaGE database. The first experiment aims to verify the integrity of the database by evaluating the performance of estimated RTFs and comparing them to their simulated counterparts \cite{habets2006room}. In the second experiment, we analyze the performance of the method from \cite{talmon2013relative} for supervised enhancement of noisy RTF estimates.

%To demonstrate the use of the database, we describe two  experiments. In the first experiment, we will test how the blocking ability of pre-computed RTFs changes between simulated and real conditions. In the second experiment, we analyze the performance of the method from \cite{talmon2013relative} for supervised enhancement of noisy RTF estimates.
%\subsubsection*{Definitions}
 We will give now two applications that can benefit from this database. The database can be used in many more applications.

Given a pair of microphones $i$ and $j$, their noisy observations of a directional source can be described as
\begin{equation}
    \begin{aligned}
        x_i &= s*h_i + v_i, \\
        x_j &= s*h_i*g_{i,j} + v_j,    
    \end{aligned}
\end{equation}
where $x_{i}$ is the observed signal by the $i$th microphone, $i=1,2$, $s$ is the source signal, $h_{i}$ is the room impulse response between the source and the $i$th microphone, $v_{i}$ is the noise signal in the $i$th microphone, and $g_{i,j}$ is the relative impulse response (ReIR) between microphones $i$ and $j$ (reference). It holds that $h_j=h_i*g_{i,j}$. 

The frequency domain counterpart of the ReIR is the Relative Transfer Function (RTF), a term which we will use in the following as it is more frequently used in the literature \cite{gannot2001}. An ATF could be seen as a special case of RTF where the non-reference microphone is a virtual one whose output is the original (non-spatial) signal $s$ (then, $h_i=\delta$ and $g_{i,j}=h_j$). Alternatively, this virtual microphone can be substituted by a microphone that is very close to the source (e.g., microphone 31 in our database).

%\INSsg{The context is unclear. You should start by mentioning the GSC. Otherwise it is unclear why you prefer to attenuate the signal and not the noise. Also discuss, why important for robustness to avoid desired source leakage into the NC branch.}
RTFs are commonly used to construct beamformers, e.g. as a blocking matrix in the Generalized Sidelobe Canceller (GSC) implementation. Higher attenuation of the desired signal (i.e. less signal leakage after blocking) is related to lower distortion at the GSC output \cite{gannot2001}. We can therefore use the blocking ability of an RTF as an objective measure of quality. Given an RTF $g$, we define {\em the blocking ability of $g$} for microphones $i$ and $j$ as
\begin{equation}
    %BA = -10 log_{10} \left( \frac{\frac{\sum s_{r}^2}{\sum v_{r}^2}}{\frac{\sum s^2}{\sum v^2}}     \right)
    {\rm BA}_{i,j}(g) = -10 \log_{10} \left( \frac{{\rm var}[s_{r}] / {\rm var}[v_{r}]}{{\rm var}[s]/{\rm var}[v]}     \right)
    \label{eq:BA}
\end{equation}
where ${\rm var}[\cdot]$ stands for the variance of the argument, $s_{r}$ is the residual of $s$ at the output of the blocking operation $x_i*g-x_j$, that is, $s_{r} =  s*h_i*g - s*h_j$, and, similarly, $v_r$ is the residual of the noise, $v_r = v_i*g - v_j$. This measure reflects the attenuation of the source relative to the attenuation of the noise. Ideally, when $g$ corresponds to the exact RTF, i.e. $g=g_{i,j}$, then $s_{r} =0$ and ${\rm BA}_{i,j}(g)=+\infty$. 

%Blocking ability of a single position in the grid can be defined as the mean of the individual RTF blocking abilities for that position.

\subsection{Experiment 1}
In this experiment, we consider grid positions [:,1,1]. The dataset recordings of the white noise for these positions observed by array 1 were taken (microphones 1 through 5). The recordings were used to compute RTFs using time-domain least squares, with microphone 1 being the reference. These positions will be referred to as reference positions.

Then, a source at position [18,1,5] (referred to as test position) is considered. Spatial images of a white noise signal played from the test position were simulated using the relevant ATFs of the database. The ATFs were obtained similarly using the database recordings for the test position by mics 1 through 5 and the original excitation signal\footnote{That is, we do not use microphone 31 here for computation of the ATFs.}. % and mic 31. 

\begin{figure}
    \centering
    \includegraphics[width=\columnwidth]{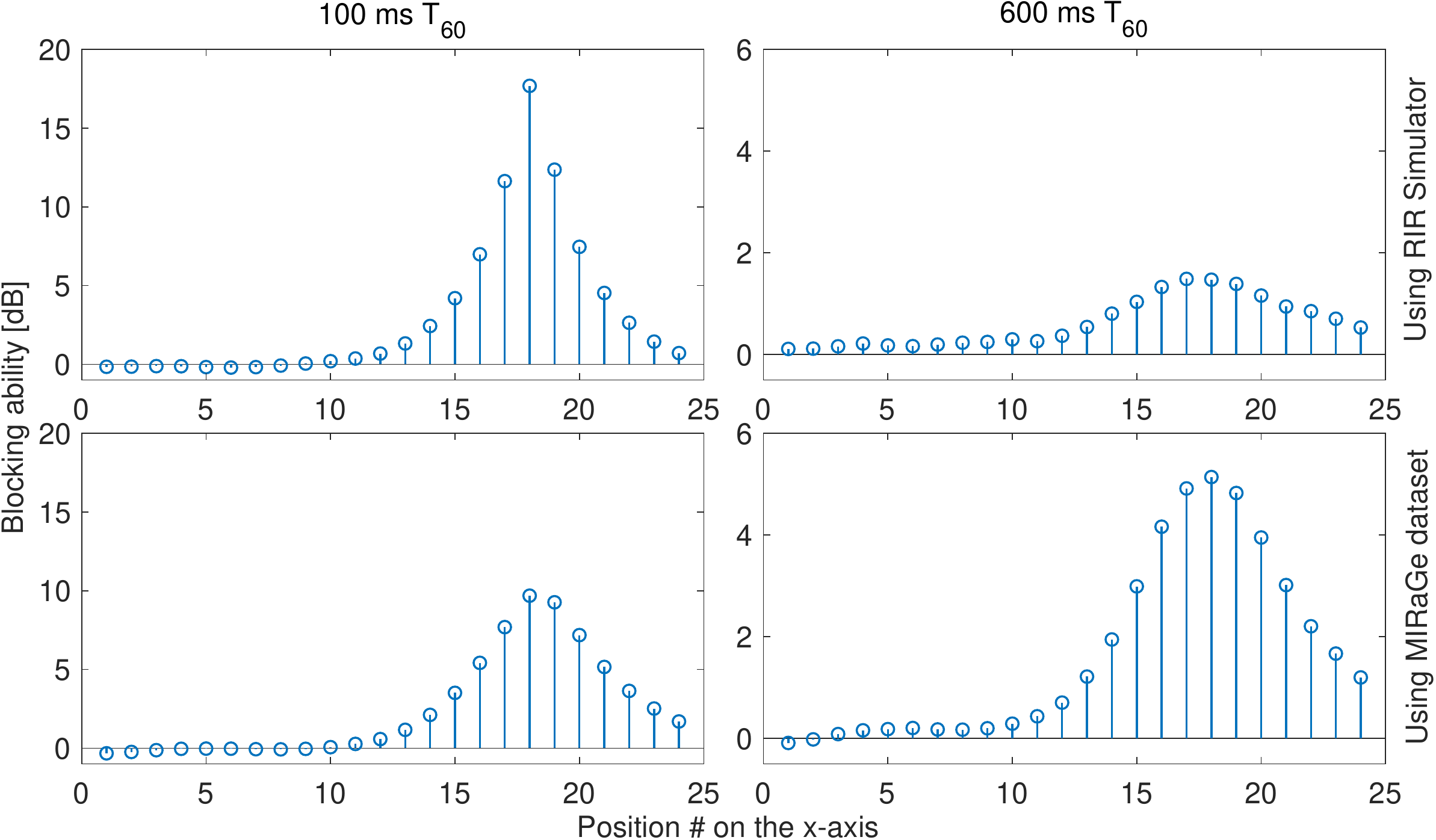}
    \caption{Blocking ability per reference positions.} % 100 ms $T_{60}$ (left), 600 ms $T_{60}$ (right, simulated setup (top) and real setup (bottom)
    \label{fig:exp1_chart}
\end{figure}

Now, the blocking ability of the RTFs when considering the source in the test position were computed. Signals were downsampled to 16~kHz. The time-domain length of the ATFs was 7168 taps and that of the RTFs was 1536 taps. $T_{60}$ levels examined were $100$~ms and $600$~ms, respectively. To compare, the entire experiment was also  simulated using the RIR generator \cite{habets2006room}. The blocking ability of the RTFs averaged over microphones 1 through 5 as a function of the reference position is shown in Fig.~\ref{fig:exp1_chart}.
% This could be seen as having a static source signal in the testing position and steering a blocking beam towards each of the reference position. We have also reversed this experiment, that is the blocking beam is steered only towards the testing position, while the source moves through each reference position.

The results show that the closer the reference position is to the test position (\#18), the higher is the blocking ability. With higher reverberation time, the overall blocking ability is lower, which is due to the insufficient time-domain length of the RTFs (1536 taps). The overall trends of the blocking ability for the real and the simulated experiment are similar, nevertheless, the values are different. This shows that the reverberation tails of the RIRs from the laboratory are significantly different from those of the simulated room.

% Varying the position of the testing point also allows us to visualize the changes in blocking ability of the reference RTFs. See our demo (footnote to GitHub) for more info.

%old version:
%In this experiment, we selected the front-most bottom row of grid positions to serve as a training set. Using time-domain least squares, we calculated RTFs from real recordings on microphones Mic 1 through Mic 5 (Array 1), with Mic 1 being the reference microphone. Then we selected a different point from the grid to serve as the source location. Using this point's ATFs, we created a spatial sound image of white noise on all microphones. Finally, we calculated the blocking ability of the training RTFs according to (equation). Sampling frequency was set to 16~kHz. Length of ATFs was 7168 taps in time-domain, length of RTFs was 1536 taps in time-domain. $T_{60}$ was set to 100 ms and 600 ms. 

%The setup of this experiment was also simulated using the RIR generator \cite{} in order to compare the results. Results can be seen in Figure \ref{fig:exp1_chart}.

%We can see there's a clear trend - the closer the training point is to the testing point, the better its blocking ability. With higher reverberation time, however, we can observe a high degradation of blocking ability, especially for the simulated setup. Reason for this?
%Varying the position of the testing point also allows us to visualize the changes in blocking ability of the training RTFs. See our demo (footnote to GitHub) for more info.

\subsection{Experiment 2}
In \cite{talmon2013relative}, a method, denoted manifold projection (MP), for supervised RTF identification was introduced for the purpose of increasing the robustness of optimal beamformers. In this method, a manifold of typical RTFs in a particular room is learned in advance and then exploited to improve the identification of unknown RTFs based on noisy measurements. The method was evaluated in \cite{talmon2013relative} only using a simulated experiment and was shown to provide superior blocking ability over the noisy measurement-based RTF, especially in low SNR conditions. In our experiments, we evaluate the performance of this method in realistic conditions using our database.

First, we repeated the experimental setup in \cite{talmon2013relative} as closely as possible with our database. We used every second position in the grid's $x$-axis, every position of the $z$-axis and the first position of the $y$-axis (i.e. [1:2:24,1,:], for a total of $12\times 1\times 9$ positions). For each of these positions, the RTF was calculated between microphones 14 and 15 (Array 3, $8$~cm inter-mic distance, $3$~m in front of the grid) using clean excitation white noise signals. These RTFs serve as the training set in the supervised methods. The same setup was also created using the RIR simulator.

\begin{figure}[t]
    \centering
    \includegraphics[width=\columnwidth]{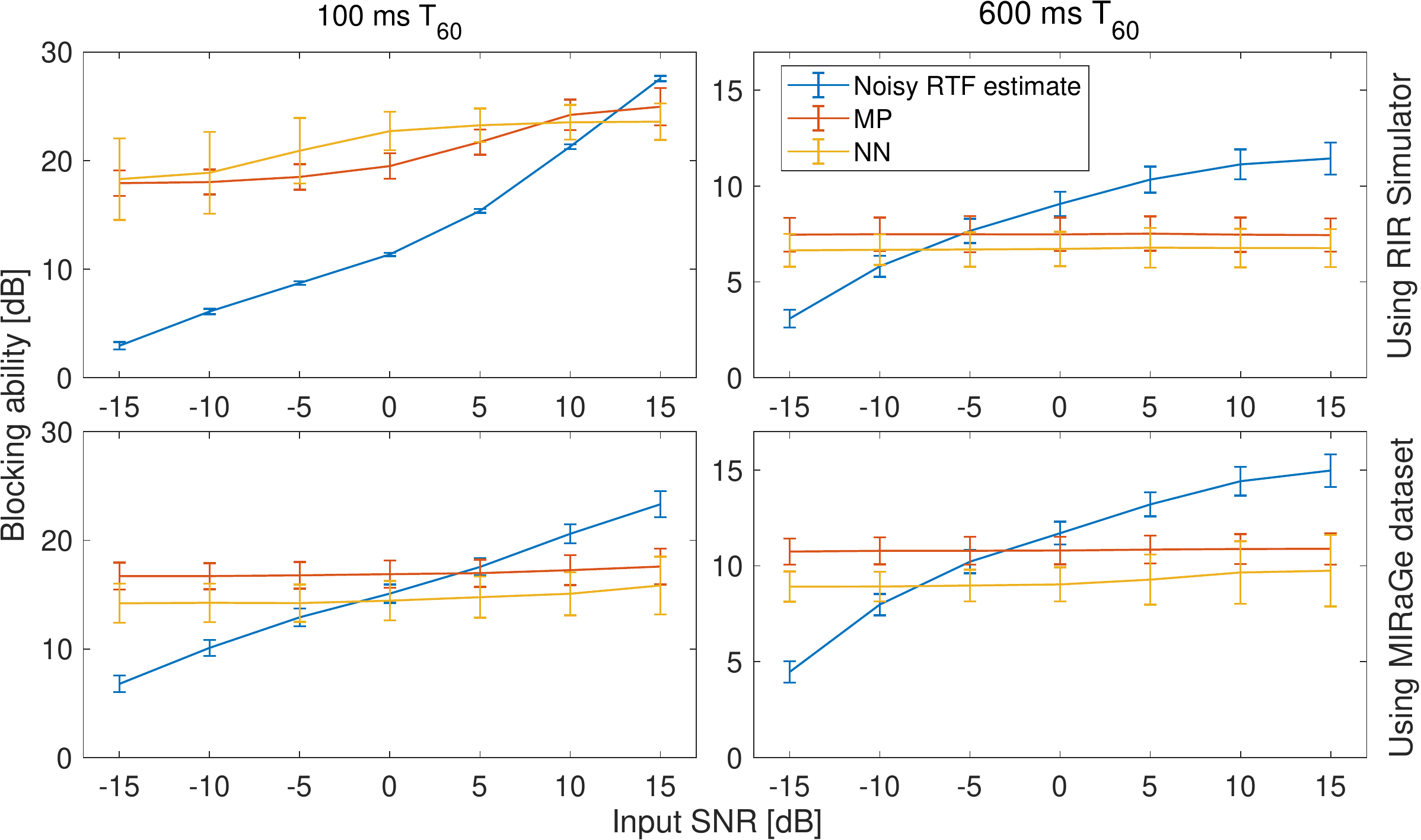}
    \caption{Results of the second experiment} %, 100 ms $T_{60}$ (left), 600 ms $T_{60}$ (right), simulated setup (top) and real setup (bottom)
    \label{fig:exp2_1000_iters}
\end{figure}

In one trial of the experiment, a new position within the grid except any positions in [:,1,:] is randomly selected. %Our database is limited by a finite number of pre-recorded positions ($216$ in the selected plane, $108$ of which are training positions), however, and this low variability would affect the experimental results. 
%We have therefore decided to use the rest of the database in its entirety (all of the remaining vertical planes, totalling 3888 possible positions) for testing purposes.
A 3-second long speech signal is simulated from that position. An uncorrelated spatial white noise at different signal-to-noise ratio (SNR) levels is added. From the noisy mixture, an RTF is estimated using the frequency-domain estimator proposed by Gannot et al. \cite{gannot2001}. This noisy RTF estimate is then compared to the training RTFs using the Euclidian distance, and the closest RTF is chosen in place of the noisy one. This approach is referred to as Nearest Neighbor (NN). To compare, the noisy RTF estimate is also enhanced using MP \cite{talmon2013relative}. The resulting RTF estimates were evaluated using the blocking ability with respect to the simulated source. The other parameters were the same as in the first experiment. The average results of 1000 independent trials are shown in Fig.~\ref{fig:exp2_1000_iters}.

The results show that both supervised methods provide superior blocking ability compared to the unsupervised method (the noisy RTF estimate) when the input SNR is lower than $0$~dB or $-10$~dB, depending on the reverberation time. While both NN and MP methods perform similarly in the simulated conditions and low reverberation time, MP provides consistently better results than NN in terms of the blocking ability when using real data. %\INSsg{Is Fig. 6 with real database or simulated? It is unclear. Also, it seems that noisy is better for high T60, in a wider range of SNRs. This is a bit surprising.}

\section{Conclusions}
We have introduced a new database of dense measurements within a 3D area of an acoustic lab.
We have shown that the database can be used for detailed analyses of spatial processing algorithms subject to source location within the measured area. Small variations of RIRs and ReIRs (resp. ATFs and RTFs) due to small changes of the source location can be observed. The database provides an alternative to the popular room impulse response simulator based on the image method \cite{allen1979,habets2006room}. %, which simulates an artificial room, while the database comes from a real-world room. 

% -------------------------------------------------------------------------
% Either list references using the bibliography style file IEEEtran.bst
% \balance
%\bibliographystyle{IEEEtran}

%\bibliography{refs19,ISI}
%
% or list them by yourself
% \begin{thebibliography}{9}
% 
% \bibitem{waspaa19web}
%   \url{http://www.waspaa.com}.
%
% \bibitem{IEEEPDFSpec}
%   {PDF} specification for {IEEE} {X}plore$^{\textregistered}$,
%   \url{http://www.ieee.org/portal/cms_docs/pubs/confstandards/pdfs/IEEE-PDF-SpecV401.pdf}.
%
% \bibitem{PDFOpenSourceTools}
%   Creating high resolution {PDF} files for book production with 
%   open source tools, 
%   \url{http://www.grassbook.org/neteler/highres_pdf.html}.
%
% \bibitem{eWilliams1999}
% E. Williams, \emph{Fourier Acoustics: Sound Radiation and Nearfield Acoustic
%   Holography}. London, UK: Academic Press, 1999.
% 
% \bibitem{ieeecopyright}
%   \url{http://www.ieee.org/web/publications/rights/copyrightmain.html}.
%
% \bibitem{cJones2003}
% C. Jones, A. Smith, and E. Roberts, ``A sample paper in conference
%   proceedings,'' in \emph{Proc. IEEE ICASSP}, vol. II, 2003, pp. 803--806.
% 
% \bibitem{aSmith2000}
% A. Smith, C. Jones, and E. Roberts, ``A sample paper in journals,'' 
%   \emph{IEEE Trans. Signal Process.}, vol. 62, pp. 291--294, Jan. 2000.
% 
% \end{thebibliography}
\section*{Appendix: MIRaGe utilities screenshots}

\begin{figure}[H]
    \centering
    \includegraphics[width=\columnwidth]{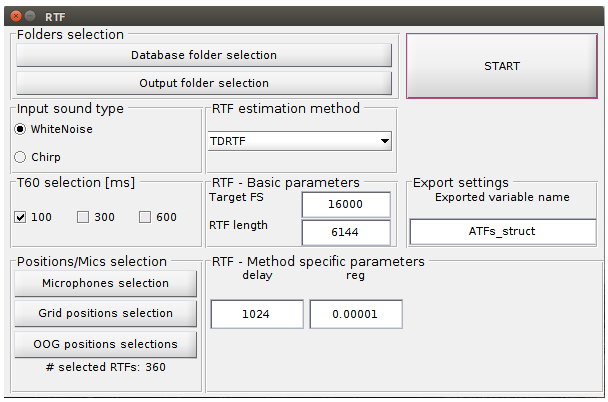}
    \caption{Main window}% 100 ms $T_{60}$ (left), 600 ms $T_{60}$ (right, simulated setup (top) and real setup (bottom)
    \label{fig:utils_main_window}
\end{figure}

\begin{figure}[H]
    \centering
    \includegraphics[width=\columnwidth]{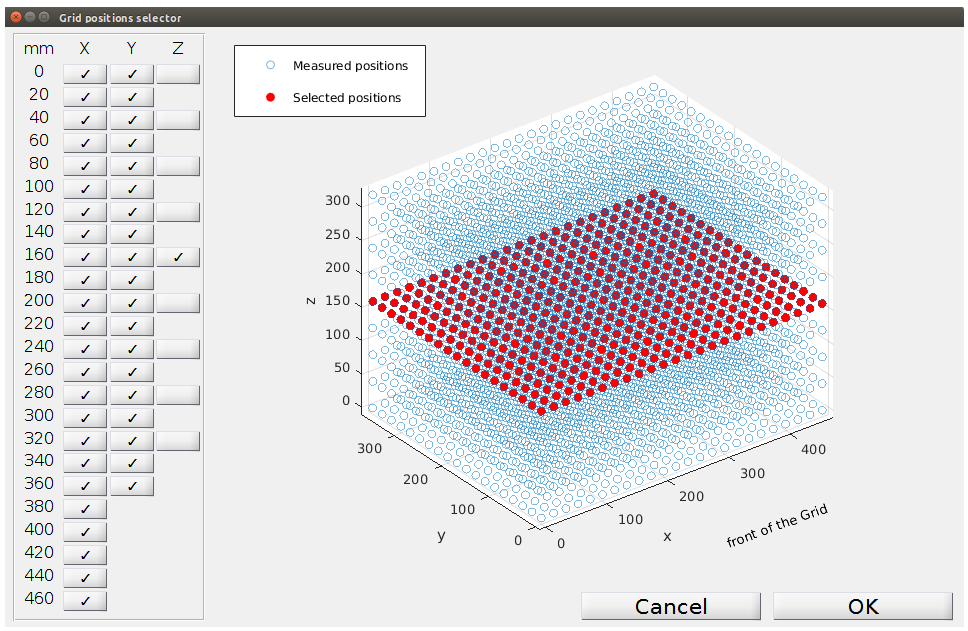}
    \caption{Grid positions selection window} % 100 ms $T_{60}$ (left), 600 ms $T_{60}$ (right, simulated setup (top) and real setup (bottom)
    \label{fig:utils_grid_window}
\end{figure}

\begin{figure}[H]
    \centering
    \includegraphics[width=\columnwidth]{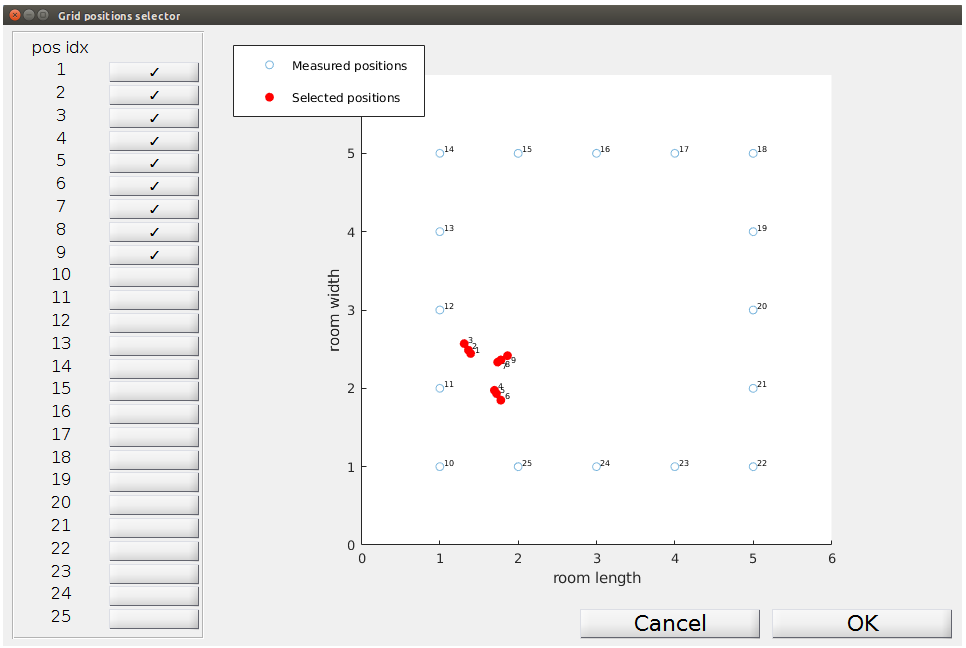}
    \caption{Out-Of-Grid positions selection window} % 100 ms $T_{60}$ (left), 600 ms $T_{60}$ (right, simulated setup (top) and real setup (bottom)
    \label{fig:utils_oog_window}
\end{figure}

\begin{figure}[H]
    \centering
    \includegraphics[width=\columnwidth]{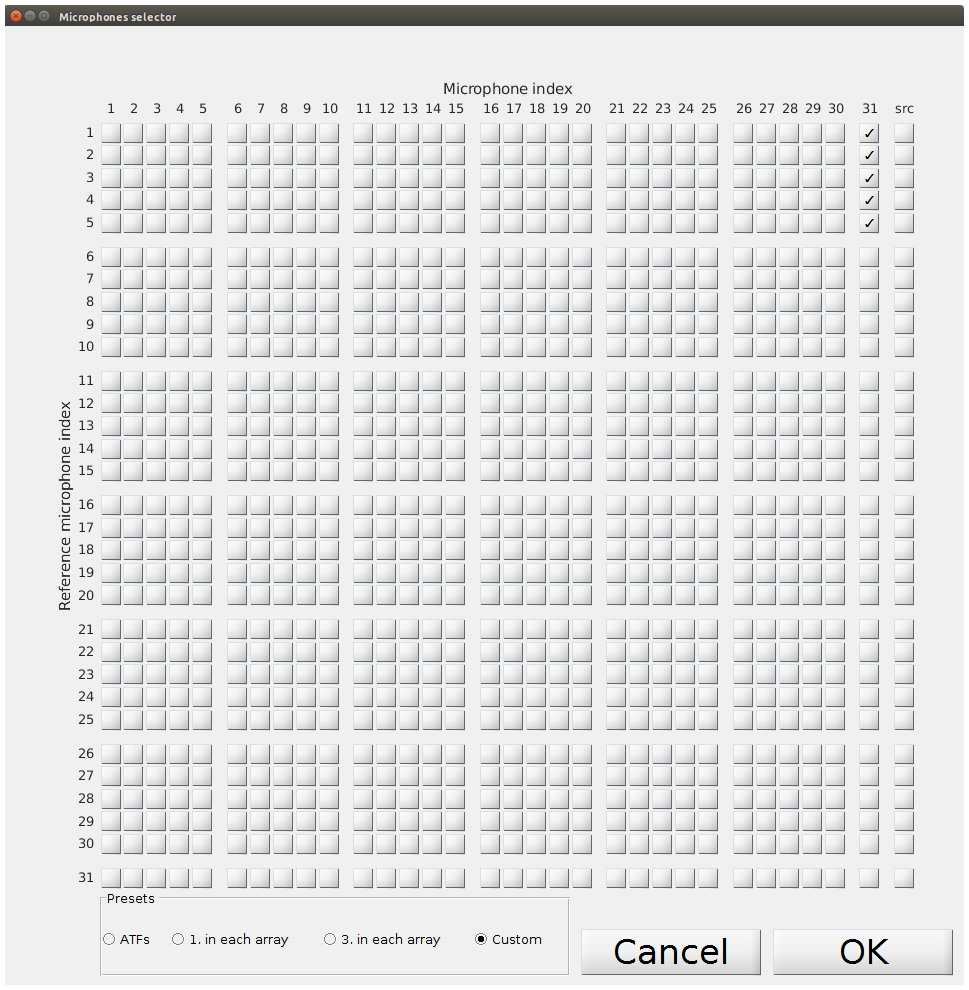}
    \caption{Microphones selection window} % 100 ms $T_{60}$ (left), 600 ms $T_{60}$ (right, simulated setup (top) and real setup (bottom)
    \label{fig:utils_mics_window}
\end{figure}

\end{document}